\documentclass[a4paper,11pt]{article}
\pdfoutput=1
\usepackage{jcappub}
\usepackage[T1]{fontenc}
\usepackage{xcolor}
\usepackage{gensymb}
\usepackage{amsmath}
\usepackage{array}

\newcolumntype{C}[1]{>{\centering\let\newline\\\arraybackslash}m{#1}}

\title{Addressing the self-interaction for ELDER dark matter from the 21-cm signal}
\author[a]{Rupa Basu}
\author[b,c]{Debasish Majumdar}
\author[d]{Ashadul Halder}
\author[a]{Shibaji Banerjee}

\affiliation[a]{St. Xavier's College (Autonomous), 30 Mother Teresa Sarani, Kolkata-700016}
\affiliation[b]{Theory Division, Saha Institute of Nuclear Physics, 1/AF Bidhannagar, Kolkata 700064, India}
\affiliation[c]{Homi Bhabha National Institute, Anushakti Nagar, Mumbai 400094}
\affiliation[d]{S. N. Bose National Centre for Basic Sciences, JD Block, Sector III, Salt lake city, Kolkata-700106, India}

\emailAdd{rupabasu.in@gmail.com}
\emailAdd{debasish.majumdar@saha.ac.in}
\emailAdd{ashadul.halder@gmail.com}
\emailAdd{shiva@sxccal.edu}

\abstract{The self-interacting dark matter can affect various cosmological processes. Such interactions can be number conserving (\emph{e.g.} $2 \rightarrow 2$) or number violating (\emph{e.g.} $3 \rightarrow 2,\,4 \rightarrow 2$ etc.). The latter processes where three (or more) dark matter particles undergo self-annihilation/scattering to produce less number of dark matter is termed as ``Cannibalism'' process. In this work, the self-interaction of dark matter and the strength of such interactions are investigated in the light of experimental results of the global 21-cm spectrum of neural hydrogen from the era of cosmic dawn. From the present work, it appears that $2\rightarrow 2$ process is much more dominant over the $3\rightarrow 2$ process. It is also found that such interactions affect the dark matter-baryon elastic scattering cross-section. The study also indicates the presence of multi component dark matter of different mass range in the Universe.}

\begin{document}
	\maketitle
	\flushbottom
	
\section{Introduction}

A strong absorption signal of the global 21-cm hyperfine transition  spectrum of neutral hydrogen around redshift $z =17.2 $ has been observed by the EDGES experiment {\cite{edges}}. EDGES reported an absorption spectrum centred at a frequency of $\sim 78.2$ MHz which corresponds to a redshift of $\sim 17.2$. The result reported by EDGES experiments was estimated to be $3.8\,\sigma$ off from the prediction in the standard $\Lambda$CDM model.

At the epoch of cosmic dawn $(15\leq z \leq 35)$ \cite{cosmicdawn}, in particular when the first stars were born at about $z \sim 20$ the cosmic gas hydrogen gas was at its coolest point before being heated up by the X-ray radiation. This ionizes the neutral hydrogen and heralds the reionization epoch ending the ``Dark Ages''. The first stars are born and the Universe has been lit up. The 21-cm hydrogen line is emitted due to the hyperfine transition between the ortho (the spins of proton and electrons are parallel) and para states (the spins are anti-parallel) of hydrogen at its ground state. At the era of reionization after the first star was born the spin temperature $T_s$ of hydrogen gas is coupled to the cooler gas and at that epoch $T_s < T_{\rm{CMB}}$ thereby absorption line of 21-cm would be observed at the CMB radio background. Thus depending on whether the spin temperature is greater or less than the background radio signal temperature, the emission or absorption line of the 21-cm H1 spectrum can be observed. The anomaly between the experimental and standard model prediction of the brightness temperature of the  redshifted 21-cm hydrogen absorption spectra has inspired several probes in the dark mystery of the early universe, in particular during the cosmic dark ages. As the known Universe is made up of $75 \%$ hydrogen gas, so the 21-cm transition measurement could be significant to understand the evolution of the Universe as well as the evolution of the temperature of the hydrogen gas. Its intensity depends on the population ratio of the triplet and singlet state. This is expressed in terms of a quantity called spin temperature or $T_s$. Since the 21-cm signal can be measured with respect to the cosmic microwave background radiation or CMB temperature ($T_{\rm{CMB}}$), the 21-cm brightness temperature $T_{21}$ is expressed in terms of $T_s$ and $T_{\rm{CMB}}$ and optical depth $\tau$ of the medium through which the radiation passes. It may be noted that the process of reionization increases the number density of free electron which could scatter off CMB affecting the optical depth $\tau$.

At $z \sim 200$, the cosmic gas decouples from the Cosmic Microwave Background (CMB) and its temperature drops below radiation due to the adiabatic evolution of the gas temperature. 
The unexpected cooling of the 21-cm signal observed by the EDGES experiment has turned out to be a promising probe to understand the cosmic history during the cosmic dawn era. The influence of baryon-dark matter interaction as also the effect of dark matter decay on the evolution during the cosmic dawn era can also be addressed by studying the 21-cm signal from that era. The  global 21-cm signal detected by EDGES from the era of the cosmic dawn appears to be at a temperature considerably lower than that expected from the known standard cosmological model ($\Lambda$CDM). 

The apparent cooling of the gas reported by the EDGES experiments could have been  caused by a process in which heat is absorbed from baryons. In case the heat exchange between dark matter and baryons causes this observed cooling effect then it can be envisaged that the scattering between the dark matter and baryons could lead to this effect. For this to happen however the mass of dark matter should be less or of the order of the mass of baryons. The dark matter annihilation also can cause the exchange of heat with the surroundings. In the present work, a sub-GeV dark matter is considered. The scattering of this sub-GeV dark matter with baryons could have a significant cooling of the baryon fluid once Ly-$\alpha$ radiation becomes effective \cite{rennan_3GeV}.

A strong velocity-dependent DM-baryon (here baryon refers to all standard model particles) scattering would lead to a strong 21-cm signal at the cosmic dark ages and also elude the astrophysical and cosmological constraints at the present day. However, for  Rutherford-like DM scattering with baryons (comparable to Compton scatterings), the interaction couples to the gas with CMB radiation. In this case, the scattering cross-section has large velocity dependence. The DM-baryon elastic scattering around the reionization epoch may cause the gas to cool down further causing a lower brightness temperature of 21-cm absorption line.  The nature of velocity-dependent scattering cross-section is generally expressed as $\sigma = \sigma_0 \, v^{n}$ \cite{dvorkin2020cosmology, Nadler_2019, Bhoonah_2018, Kovetz_2018, Mack_2007} where $n$ is a constant that depends on the nature of dark matter. For example, in the literature, it has been discussed that for magnetized DM, the index $n= \pm 2$, while for millicharged dark matter $n=-4$ \cite{mcharge1,mcharge2, Aboubrahim:2021ohe, PhysRevD.98.103005, PhysRevLett.121.011102, PhysRevD.98.023013}. The $\,v^{-4}(n=-4)$ dependence of the DM-baryon scattering, is also true for Rutherford-like DM-baryon scattering \cite{munoz}. If the scattering potential is Yukawa type \cite{yukawa} then the parameter $n$ in the scattering cross-section $\sigma = \sigma_0 v^n$ can take the value $n=2,1,0,-1$. In addition to DM-SM scattering the DM can also undergo the self-scattering processes. The DM-DM self-scattering may also influence the brightness temperature of the 21-cm absorption line.  
In the present work, in addition to the {\it{(i)}} DM-baryon elastic scattering ($\chi + \rm{SM} \rightarrow \chi + \rm{SM}$) and {\it{(ii)}} DM annihilation to SM ($\chi + \chi \rightarrow \rm{SM + SM}$) ( $\chi \chi\chi \rightarrow \chi\chi$)($\chi$ represents the dark matter), the heating (cooling) effects due to various dark matter self-scattering processes are also addressed. Here two types of self-scattering are considered. These are {\it{(i)}} $``2 \rightarrow 2" $ elastic self-scattering ($\chi \chi \rightarrow \chi\chi$) and {\it{(ii)}} $``3 \rightarrow 2" $
self-annihilation {\cite{elder, elderphenomenology}}.

The self-interacting DM is a promising probe to the dark sector of the Universe under the uncharged SM gauge group. If the relic density of DM is determined by either the cross-section of the elastic scattering between DM and SM or the cross-section of the number-changing self-interaction processes, then the scenario is called ``Elastically Decoupling Relic'' or ELDER. In this case, the mass of DM $(m_\chi) \sim$ the QCD confinement scale ($(10 {\rm{MeV}} < m_\chi < 100 {\rm{MeV}})$). In this work,  all four interactions mentioned above are assumed to be effective at early universe ($\Gamma > H$, where $\Gamma$ is the respective interaction rate and $H$ is the Hubble paraneter) until ELDER dark matter becomes non-relativistic. Under these circumstances the interaction rates for the process $\chi + \rm{SM} \rightarrow \chi + \rm{SM}, \,\, \chi \chi \rightarrow \chi\chi \,\,{\rm{and}} \,\,\chi \chi\chi \rightarrow \chi\chi$ fall off exponentially whereas elastic scattering interaction ($\chi + \rm{SM} \rightarrow \chi + \rm{SM}$) rate varies more slowly.  

In the present work, the principal focus is on these interactions of ELDER dark matter on the basis of 21-cm observational results. The experimental value of the brightness temperature of 21-cm signal {\textit{i.e}} $T_{21} = -500_{-500}^{+200}$ mK at the redshift $z=17.2$ has been adopted to constrain the couplings of ELDER dark matter interactions. It may be mentioned that in another recent work related to SARAS experiment \cite{saras} regarding a global 21-cm signal, the EDGES results have been challenged. However, in this work we adopted only the EDGES results.

This paper is organised as follows. Sec.~\ref{sec:2} gives a brief account of ELDER dark matter model. In Sec.~\ref{sec:3}, the thermal evolution of the 21-cm signal has been discussed. The cooling/heating of gas through the four active reactions of ELDER dark matter model is addressed in Sec.~\ref{sec:4}. The analysis and the results in this work are given in Sec.~\ref{sec:5}. Finally, in Sec.~\ref{sec:6} some concluding remarks along with the summary are given.

\section{\label{sec:2}Elder dark matter model and its  cannibalism property}
 Although the existence of DM in the Universe is well established but its particle nature and interaction properties are yet to be understood. Popular candidates of dark matter include ``weakly interacting massive particles'' or ``WIMPs'', the particle nature of which is generally formulated using theories beyond the standard model (SM) such as super-symmetry, theories of extra dimension or formulating theories by  simple extensions of the standard model. Such WIMPs are generally thermal and relic density is obtained by the process of thermal decoupling. The unknown dark matter can also be classified according to their possible processes of decoupling from the Universe's SM plasma. Thus in addition to the ``freeze-out'' cold DM (\textit{e.g,} WIMPs) there could be ``freeze-in'' DM (particle density approaches the equilibrium density) etc. But for these types of DM, mainly the annihilation process $\chi + \chi \rightarrow {\rm{SM}}+ {\rm{SM}}$ dominates the process of decoupling and relic abundance. On the other hand, where other processes (along with $\chi + \chi \rightarrow {\rm{SM}}+ {\rm{SM}}$) such as elastic scattering ($\chi + {\rm{SM}} \rightarrow \chi + {\rm{SM}}$) and self-annihilation are also involved in the decoupling of dark matter then new scenario for the type of dark matter emerges. To this end, the freezing of the comoving number density of DM is assumed to be guided by the four processes namely 
 
 \begin{itemize}
     \item  DM annihilation to SM ( $\chi + \chi \leftrightarrow \rm{SM + SM}$)
	\item  DM-baryon elastic scattering ($\chi + \rm{SM} \leftrightarrow \chi + \rm{SM}$) 
	\item $``3 \rightarrow 2" $
	self-annihilation( $\chi \chi\chi \leftrightarrow \chi\chi$)  
	\item  $``2 \rightarrow 2" $ elastic self-scattering ($\chi \chi \leftrightarrow \chi\chi$). 
\end{itemize}

 In a proposed scenario, the variation of DM particle number and number density involve the four processes mentioned above but the relic density is determined by the elastic scattering process. Such a type of DM with  an elastically decoupling relic is termed as ELDER (elastically decouple relic) DM. In this scenario, when the temperature of the Universe, $T_{\rm{Univ}} < m_\chi$ (mass of the DM) the equilibrium density of $\chi$ drops exponentially as $e^{-\frac{m_\chi}{T}}$, $T$ is the temperature of SM plasma, leading to the decoupling of the annihilation process. But the heat generated due to self-annihilation and elastic scattering process still keeps the dark matter $\chi$ in thermal equilibrium with the SM. In this present ELDER scenario although elastic scattering decouples first, self-annihilation processes ($\chi\chi \chi \rightarrow \chi\chi$ and $\chi \chi \rightarrow \chi\chi$) keep the temperature of the DM almost constant even though the Universe is expanding. The freezing of the comoving number density of ELDER DM $\chi$ occurs with the decoupling of the self-annihilation process. It may be noted that the relic abundance of ELDER DM remains almost the same as the DM density during the decoupling epoch of elastic scattering. It may be mentioned here that in the self-annihilation process $\chi \chi \chi \rightarrow \chi \chi$, three DM particles are reduced to two DM particles. As mentioned earlier, this is what is termed as ``cannibalisation'' or ``cannibalism'' of DM \cite{elder, elderphenomenology,cannibal}.  In this process however, the missing particle manifests itself as the energy which is carried away by the outgoing particles. Also the process $\chi \chi \rightarrow \chi \chi $ may be rapid and heat produced from all these self-annihilation processes keep the cannibalising particle at almost constant temperature with vanishing chemical potential. It is also to be noted that the $``3 \rightarrow 2"$ cannibalism process decouples earlier than the $``2 \rightarrow2"$ self-annihilation process. This is because the self annihilation rates $\Gamma_{33} \propto  n_\chi^2$ for $``3 \rightarrow 2"$ process, while that for  $``2 \rightarrow 2"$ process, $\Gamma_{22} \propto  n_\chi$. 

This may now be mentioned from the above discussions that the relic density of such cannibal ELDER DM depends on the cross-section of self-interaction processes and the cross-section of the elastic scattering process. This can be naturally obtained if $m_\chi \sim$ QCD confinement scale ($10 - 100$ MeV)  \cite{elder, elderphenomenology}. 
The relic abundance of ELDER dark matter of mass range $10$ MeV to $100$ MeV is given by  {\cite{elder}}

\begin{equation}
	\Omega_\chi \sim \dfrac{10^6\, m_{\rm{MeV}}\,{\rm{exp}}(-10\, \epsilon_{-9}^{1/2}\, {m_{\rm{MeV}}^{-1/4})}}{1+0.07 {\rm{log}}\alpha}.
	\label{omegax}
\end{equation}
The quantities $\alpha$ and $\epsilon_{-9}( =\epsilon/10^{-9})$ are the coupling strength of $``3 \rightarrow 2"$ self-annihilation and $``2 \rightarrow 2"$ self-scattering respectively and $m_{\rm{MeV}} = \dfrac{m_\chi}{1 {\rm{MeV}}}$ is the parameterized mass of dark matter in MeV. The recent works show that for ELDER dark matter, the coupling strength $\alpha$ should lie in the range   $0.5 \dfrac{m_\chi}{10{\rm{MeV}}}\leq\alpha\leq 73$ {\cite{elderphenomenology}}.




\section{\label{sec:3}Thermal evolution of 21-cm signal}

The origin of 21-cm signal due to the transition of electrons between two spin states ($s=0$ and $s=1$ spin state) of the neutral hydrogen atom. The intensity of this hydrogen absorption spectra is described using the brightness temperature $T_{21}$ and is given by \cite{Pritchard_2012,munoz}

\begin{equation}
	T_{21} =  \frac {T_s - T_\gamma} {1+z} (1 - e^{-\tau}) .
	\label{t21}
\end{equation}
In the above expression, $\tau$ represents the optical depth of the medium, $T_{\gamma}$ is the background temperature and $T_s$ is the spin temperature at redshift $z$. The spin temperature $T_s$ can be expressed as the ratio of the number densities of hydrogen atoms having two different spin states $n_0$ and $n_1$ (number density $n_0$ and $n_1$ corresponding to the $s=0$ and $s=1$ states respectively), given by \cite{Pritchard_2012,munoz}
\begin{equation}
	\dfrac{n_1}{n_0} = \frac {g_1}{g_0} \exp (-h\nu / k \,T_s) = 3\exp (-T_*/ T_s) \,\,,
	\label{boltzmann}
\end{equation}  
where the factors $g_0$ and $g_1$ are the degeneracy factors for spin states $s=1$ and $s=0$ respectively and $k$ is the Boltzmann's constant. In the above expression, $h \nu =E_{0} = 5.9 \times 10^{-6} {\rm{eV}}$ is the energy difference of the hyperfine splitting of a Hydrogen atom at the ground state, where $\nu$ and $T_*$ are the frequency and temperature of the 21-cm spectrum respectively.

The evolution of the spin temperature $T_s$ essentially depends on three factors in addition to the Hubble parameter $H(z)$ as described below.
\begin{itemize}
	\item The collision of $H-H$ and $H-e$ influences in the transition between the spin states $s=1$ and $s=0$. This can be measured by the collisional transition rate $C_{10}$ \cite{c10,widmark201921}.
	\item The cosmic ray can be absorbed or emitted by or from cosmic hydrogen. This can be evaluated from the emission and absorption rates $B_{10}$ and $B_{01}$ and $B_{10} =\frac{B_{01}}{3}$. These absorption is related to the Einstein coefficient $A_{10}$ for spontaneous emission and  $B_{10} = A_{10} \frac{T_\gamma}{E_0} = 2.9\times 10^{-15} {\rm{sec^{-1}}}$. ($B_{10}$ is related to the transition from $s=1$ to $s=0$ while $B_{01}$ denotes $s=0$ to $s=1$ transition ratio.
	\item The emitted Lyman$\alpha$ (Ly$\alpha$) photons during the birth of first stars $(z\leq 25)$ produce Wouthuysen-Field effect by which the gas can be coupled to $T_s$ \cite{salpha,roy2009wouthuysen}. Note that after the recombination, CMB photons influence the spin temperature and $T_s \sim T_\gamma$. But due to Wouthuysen-Field effect \cite{salpha, jalpha, Kuhlen_2006, Pritchard_2012}, Ly$\alpha$ photons from newborn stars drive the spin temperature to be nearly equal to the baryon temperature $T_b$ around $z\leq 20$. 
\end{itemize}

Therefore the spin temperature $T_s$ in equilibrium can be written as 
\begin{equation}
	T_s = \frac{T_{\gamma} + y_c T_k + y_\alpha T_\alpha}{1+y_c+y_\alpha}
	\label{eq:Ts} \,\, .
\end{equation}
In the above equation, the quantities $y_c =\frac{C_{10} T_{\star}}{A_{10} T_k}$ and $y_\alpha =\frac{P_{10} T_{\star}}{A_{10} T_\alpha}$  are the collisional and Lyman$\alpha$ coupling coefficients  \cite{pritchard08,Pritchard_2012,chen} respectively. In Eq.~\ref{eq:Ts}, $T_k$ is the kinetic temperature and $P_{10}$ is the indirect de-excitation rate of hyperfine structure levels \cite{Pritchard_2012,pritchard2006descending,widmark201921}.\linebreak

The optical depth $(\tau)$ of the 21-cm signal in the medium as mentioned in Eq.~\ref{t21} also depends on the spin temperature $T_s$, the gradient of peculiar velocity $\delta_r v_r$ and the number density of neutral hydrogen $n_{\rm{HI}}$. The expression for optical depth $\tau$ is given by \cite{Pritchard_2012,munoz}
\begin{equation}
	\tau = \frac {3} {32} \frac {T_*}{T_s} n_{\rm HI} \lambda_{21}^3 \frac {A_{10}}
	{H(z)+(1+z) \delta_r v_r} \,\,.
\end{equation}
Considering  the simplest scenario $(T_s = T_b)$, the brightness temperature of 21-cm signal at $z=17.2$  (henceforth denoted as $T_{21}^{Th}$) is obtained as (using Eq.~\ref{t21})
\begin{equation}
	T_{21}^{Th}\sim -200 \,{\rm{mK}} \,\,.
	\label{eq:t21theo}
\end{equation}
But the observed result is far from what is given in Eq.~\ref{eq:t21theo}. The EDGES result for brightness temperature $T_{21}^{EDGES}$ at that redshift is given as \cite{edges}
\begin{equation}
	T_{21}^{EDGES} = -500_{-500}^{+200}\, {\rm{mK}} \,\,,
\end{equation}
with $99\%$ confidence level. As mentioned earlier, the inconsistency between the observed and theoretical results is almost 3.8$\sigma$.

\section{Effect of ELDER dark matter on 21-cm line}
\label{sec:4}
The brightness temperature  $T_{21}$ is modified if  one of the background temperatures $T_\gamma$ or spin temperature $T_s$ or both vary. At $z \sim 1100$ the photons should free stream but due to the number of dense photons, radiation remains coupled to matter until about $z \sim 200$ via Compton scattering. When the Compton heating timescale supersedes Hubble time, the decoupling of matter and radiation takes place and after that epoch, the matter and dark matter  temperature fall faster ($T_{m/\chi} \sim (1+z)^2$) than radiation temperature $(T_\gamma \sim (1 + z))$.

In absence of any interactions, the evolutions of matter SM gas and DM can respectively be written as \cite{PhysRevD.98.023501,Pritchard_2012}
\begin{equation}
	(1+z)\frac{{\rm d} T_b}{{\rm d} z} = 2 T_b + \frac{\Gamma_c}{H(z)}
	(T_b - T_\gamma)
	\label{tb_old}
\end{equation}
and 
\begin{equation}
	(1+z)\dfrac{{\rm d} T_\chi}{{\rm d} z} = 2 T_\chi \,\,,
	\label{tx_old}
\end{equation}
where $\Gamma_c $ is the Compton scattering rate.
In this work, we address the effect on $T_{21}$ in case the four ELDER dark matter processes (as described in Sec.~\ref{sec:2}) are included in the analysis along with DM-baryon elastic scattering and DM annihilation to SM. 

\subsection{DM-baryon elastic scattering $(\chi+SM \leftrightarrow \chi+SM)$}

In the early Universe, after the decoupling of the photons, the dark matter and baryons developed a relative velocity $V_{\chi b}$. This generates a dragging effect between the baryon and dark matter. Also, it is likely that DM-baryon undergoes Rutherford like elastic scattering whereby the scattering cross-section is strongly velocity dependent. In the present work $v^{-4}$ dependence ($v$ being the dark matter velocity) for the scattering cross-section has been adopted such that 
\begin{equation}
	\sigma = \sigma_0 v^{-4} = (\sigma_{41} \times 10^{-41} v^{-4})\, {\rm{cm^2}} \,\,.
	\label{sigma}
\end{equation}
Here the term $\sigma_{41}$ is a dimensionless quantity given by $\sigma_{41} = \dfrac{\sigma_0}{10^{-41}}$. The relative velocity $V_{\chi b}$ or drag term is the difference between the bulk DM velocity $V_\chi$ and baryon velocity $V_b$, \textit{i.e,} $V_{\chi b} = V_\chi - V_b $. In our analysis we compute the evolution of relative velocity $V_{\chi b}$ with initial condition $V_{\chi b} = V_{\chi b 0} = 10^{-4}c$ at $z=1010$. Also the scattering of  hotter baryon/gas with colder DM (in the present case this is referred to as ``interaction {\it(i)}'' of ELDER DM processes mentioned in Sec.~\ref{sec:2}) will heat up the DM. The rate $D(V_{\chi b})$ of change of relative velocity $V_{\chi b}$, heating rate of baryons, $\dot{Q_b}$ and that of DM, $\dot{Q_\chi}$ are given by a set of equations \cite{munoz}
\begin{eqnarray}
	\dfrac{{\rm d}V_{\chi b}}{{\rm d}t} &=& -D(V_{\chi b}) = \dfrac{\rho_m \sigma_0}{m_b+m_{\chi}}\dfrac{1}{V_{\chi b}^2}F(r)\,\,,
	\label{dVdt}
	\\
	\dfrac{{\rm d} Q_b}{{\rm d} t} &=& \dot{Q_b} =\dfrac{2 m_b \rho_{\chi}\sigma_0 e^{-r^{2}/2} (T_{\chi}-T_b)}{(m_b+m_{\chi})^2\sqrt{(2\pi)} u_{th}^3}+\dfrac{\rho_{\chi}}{\rho_b+\rho_{\chi}}\dfrac{m_{\chi} m_b}{m_{\chi}+m_b} V_{\chi b} D(V_{\chi b})\,\,\
	\label{dqbdt}
	\\
	\dfrac{{\rm d} Q_{\chi}}{{\rm d} t} &=& \dot{Q_\chi} = \dfrac{2 m_{\chi} \rho_{b}\sigma_0 e^{-r^{2}/2} (T_b-T_{\chi})}{(m_b+m_{\chi})^2\sqrt{(2\pi)} u_{th}^3}+\dfrac{\rho_{b}}{\rho_b+\rho_{\chi}}\dfrac{m_{\chi} m_b}{m_{\chi}+m_b} V_{\chi b} D(V_{\chi b})\,\,,
	\label{dqxdt} 
\end{eqnarray}
where  $\rho_\chi$ and $\rho_b$  respectively are the DM density and baryon density and  $\rho_m = \rho_\chi+\rho_b$. The average masses of baryon and dark matter are $m_b$ and $m_\chi$ respectively and their corresponding temperatures are $T_b$ and $T_\chi$. The term $r$ is given by, $r=V_{\chi b}/u_{th}$, $u_{th}=\sqrt{T_b/m_b+T_{\chi}/m_{\chi}}$ and the quantity $F(r)$ is expressed as 
\begin{equation}
	F(r)={\rm erf}(r/\sqrt{2})-\sqrt{2/\pi} r e^{-r^2/2}\,\,.
\end{equation}
Eq.\ref{dVdt} can be reduced to the form
\begin{equation}
    \dfrac{{\rm d} V_{\chi b}}{{\rm d}z}=\dfrac{V_{\chi b}}{1+z}+\dfrac{D(V_{\chi b})}{(1+z)H(z)} \,\,.
	\label{eq:Vxb}
\end{equation}


\subsection{DM annihilation to SM $( \chi + \chi \leftrightarrow SM + SM)$}

The energy injection to the Universe may also be caused by the DM annihilation to SM particles along with others factors such as heating, ionization or excitation of the gas. The DM annihilation causes heating up of the gas by increasing gas temperature (baryon temperature). This temperature also depends upon the number of abundance of free electron $(x_e)$ at the period of thermal decoupling. The gas cools down adiabatically and therefore a higher value of $x_e$ would delay the CMB decoupling resulting in the increase of $T_b$. The rate of energy injection for annihilation depends on velocity averaged s-wave annihilation cross-section $\langle\sigma v\rangle$ and is expressed as \cite{BH_21cm_1, PhysRevD.98.023501}
\begin{equation}
	\left(\dfrac{{\rm d} E}{{\rm d} V {\rm d} t}\right)_{\rm{ inj}}= f_\chi^2\,\rho_{\chi,0}^2 (1+z)^6 \frac{\langle\sigma v\rangle}{m_\chi} \,\,. \label{eq:dedvdt}
\end{equation}
We consider the fraction $f_\chi$ of DM  to be $1$ and the two particles in the annihilation process are indistinguishable. The energy injected to the Universe is not the same as the deposited energy. The relationship between injection energy and deposition energy is 
\begin{equation}
	\left(\dfrac{{\rm d} E}{{\rm d} V {\rm d} t}\right)_{\rm{ dep}}= f_c(z) \left(\dfrac{{\rm d} E}{{\rm d} V {\rm d} t}\right)_{\rm{ inj}} \,\,,
\end{equation}
where $f_c(z)$ is the deposition efficiency at the redshift $z$ \cite{fcz001,fcz002,fcz003,fcz004,rupadi,BH_21cm_1}.  The superscript $c$ represents different channels namely heating, ionization or excitation through which deposition occurs. The fraction of the total energy deposited in the form of baryon heating and gas ionization are respectively $\chi_h = (1 + 2 x_e)/3$ and $\chi_i =(1 - x_e)/3$ \cite{BH_21cm_2, BH_21cm_4, PhysRevD.76.061301, Furlanetto:2006wp}.   

\subsection{$``3 \rightarrow 2"$ self-annihilation $(\chi \chi \chi \leftrightarrow \chi \chi)$}
\label{32interaction}
At high temperature, the DM particles are in thermal and chemical equilibrium with SM particles. But as the Universe expands with time, the temperature drops below the mass of DM particles. At this point the  rate of elastic scattering is no longer sufficient to maintain thermal equilibrium with SM gas and the first major ``decoupling'' event occurs.  On the other hand ``Freeze-out'' of DM happens when the rate of self-annihilation falls to an extent that the chemical equilibrium could not be obtained. However in between the two phenomena (decoupling/freeze-out), self-annihilation plays the principal role to maintain the chemical equilibrium of the DM although the SM temperature is no longer same as DM temperature. During this epoch, the DM particles may undergo number changing $`` 3 \rightarrow 2"$ self-annihilation process whereby three DM particles produce only two dark matter particles at the final state (``cannibalization''). In this process the number of DM particles decreases and at the same time the DM particles release kinetic energy which is then injected to the remaining gas. The DM-SM elastic scattering transfers this excess kinetic energy to the SM gas.
\linebreak
For our work we parameterize the $`` 3 \rightarrow 2"$ self-annihilation cross-section as \cite{elder,elderphenomenology}
\begin{equation}
	\langle \sigma_{3 \rightarrow 2} v^2 \rangle = \frac{\alpha^3}{m_\chi^5}\,\,,
	\label{sigma32v}
\end{equation}
where $m_\chi$ is the mass of DM particle and $\alpha$ is the coupling for $3\rightarrow 2$ process. The rate of excess energy/heat transfer to SM particles through the elastic scattering is given by 
\begin{equation}
	\dot{Q}_{32} \sim \Gamma_{32} v_\chi^2 T_\chi \,\,.
	\label{Q32dt}
\end{equation}
The quantities $\Gamma_{32} = n_\chi ^2 \langle \sigma_{3 \rightarrow 2} v^2 \rangle$ \cite{elderphenomenology} and $v_\chi$ are the interaction rate and velocity of DM particles respectively.
Since it is a self-interacting number changing process in the DM sector, the contribution of this phenomenon is only included in the thermal evolution of DM sector and is given by 
\begin{equation}
	\kappa_{32} = -\frac{2}{3} \frac{\dot{Q}_{32}}{(1+z) H(z) n_{\chi}}\,\,.
\end{equation}

\subsection{$``2 \rightarrow 2"$ elastic self-scattering
	$(\chi \chi \leftrightarrow \chi \chi)$}
For the process $\chi \chi \leftrightarrow \chi \chi$ ($``2 \rightarrow 2 "$ elastic self-scattering) there is no change in the dark matter particle counts. It is to be noted that $``2 \rightarrow 2 "$ elastic self-scattering decouples after $`` 3 \rightarrow 2"$ process. Thus these processes in the dark sector exchange energy with the DM particles and heat is transported to the SM gas. As mentioned above, this heat transport to the SM particle occurs through DM-SM elastic scattering and the heat transfer rate is given by 
\begin{equation}
	\dot{Q}_{22} \sim \Gamma_{22} v_\chi^2 T_\chi \,\,.
	\label{Q22dt}  
\end{equation}
The $``2 \rightarrow 2" $ self-scattering interaction rate is $\Gamma_{22} =  n_\chi\langle \sigma_{2 \rightarrow 2} v \rangle$ and the self-scattering cross-section is parameterized as \cite{elder}
\begin{equation}
	\langle \sigma_{2 \rightarrow 2} v \rangle = \frac{\epsilon^2}{m_\chi^2}\,\,.
	\label{sigma22v}
\end{equation}
In the above $\epsilon$ is the coupling parameter.
Likewise for $`` 3 \rightarrow 2"$ self-annihilation,  $``2 \rightarrow 2 "$ elastic self-scattering is happened only within the dark sector, thus this reaction only affects thermal evolution of DM sector and is given by the relation
\begin{equation}
	\zeta_{22} = -\frac{2}{3} \frac{\dot{Q}_{22}}{(1+z) H(z) n_{\chi}}\,\,.
\end{equation}
With these, the evolution equations described earlier in this Section can now be modified as (incorporating the $3 \rightarrow 2$ and $2 \rightarrow 2$ effects.) \cite{corr_equs,rupadi, 21cm_mar}
\begin{eqnarray}
	(1+z)\dfrac{{\rm d} T_\chi}{{\rm d} z} &=& 2 T_\chi - \dfrac{2}{3 H(z)}\dfrac{{\rm d} Q_{\chi}}{{\rm d} t} + \left(1+z\right)\left(\kappa_{32} + \zeta_{22} \right)
	\label{txfinal}
	\\
	(1+z)\frac{{\rm d} T_b}{{\rm d} z} &=& 2 T_b + \frac{\Gamma_c}{H(z)}
	(T_b - T_\gamma)-\frac{2}{ 3 H(z)}\dfrac{{\rm d} Q_{b}}{{\rm d} t}\\
    &&-\dfrac{1}{H(z)}  \left(\dfrac{{\rm d} E}{{\rm d} V {\rm d} t}\right)_{\rm dep} \dfrac{\chi_h}{n_H} \frac{2}{3 (1+x_e+f_{\rm He})}
	\label{tbfinal}
\end{eqnarray}
The evolution of the ionization fraction ($x_e = \frac{n_e}{n_b}$) is given by \cite{munoz, BH_21cm_2, PhysRevD.98.023501,rupadi, 21cm_jan}
\begin{equation}
	\frac{{\rm d} x_e}{{\rm d} z} = \frac{1}{(1+z)\,H(z)}\left[C_P\left(n_H \alpha_B x_e^2-4(1-x_e)\beta_B 
	e^{-\frac{3 E_0}{4 k_B T_{\gamma}}}\right)-I_{\rm heat}(z)\right]\,\,,
	\label{eq:xe}
\end{equation}
which essentially depends on the number density of hydrogen $n_H$ and free electrons $n_e$, while the Peebles-C factor ($C_P$) \cite{peeble, hyrec11} is almost unity. In the above expression (Eq.~\ref{eq:xe}), the quantities $\alpha_B$ and $\beta_B$ are the effective recombination coefficient and the effective photoionization rate respectively \cite{pequignot, seager,BH_21cm_5} and the term $I_{\rm heat}(z)$ denotes the amount of heat involved in ionizing the baryons, given by \cite{PhysRevD.98.023501, rupadi}
\begin{equation}
	I_{\rm heat}(z) = \chi_i \dfrac{1}{n_b E_0} \left(\dfrac{{\rm d} E}{{\rm d} V {\rm d} t}\right)_{\rm dep}.
	\label{eq:iheat}
\end{equation}

\section{\label{sec:5}Calculations and results}

	In the present work, we investigate several interesting aspects of `ELDER' dark matter in the context of the brightness temperature of the 21-cm scenario. As discussed in Sec.~\ref{sec:2}, ELDER dark matter essentially influences four kinds of interactions and they modify the temperature of the baryonic fluid and thus alter the brightness temperature of the 21-cm signal. Dark Matter-baryon interaction is the most significant among them, which decreases the temperature difference between the baryonic fluid and the dark matter fluid. The DM candidate considered in the present analysis may cool down the baryons via DM-SM elastic scattering. On the other hand, the other interactions of ELDER dark matter namely, DM annihilation ($\chi \chi \rightarrow {\rm{SM \, SM}}$), DM self-scattering ($\chi \chi \rightarrow \chi \chi$) and DM self-annihilation ($\chi \chi \rightarrow \chi \chi \chi$) heat up the baryons either directly ($\chi \chi \rightarrow \rm{SM \,SM}$), or indirectly ($\chi \chi \rightarrow \chi \chi$ and $\chi \chi \rightarrow \chi \chi \chi$) \cite{elder}. Now incorporating all those effects of ELDER dark matter, thermal evaluations of $T_b$ and $T_{\chi}$ are obtained by solving several coupled equations (Eqs.~\ref{eq:Vxb}, \ref{txfinal}, \ref{tbfinal} and \ref{eq:xe}) simultaneously. Note that for the present CDM candidate, the temperature of DM fluid is considered to be negligible at $z\simeq 1010$. On the other hand, at $z\simeq 1010$, $T_{b}\approx T_{\gamma}$ as baryons and radiation were tightly coupled before recombination.
	
	\begin{figure}[h]
		\centering
		\begin{tabular}{cc}
            \includegraphics[trim=0 40 0 0, clip, width=0.5\textwidth]{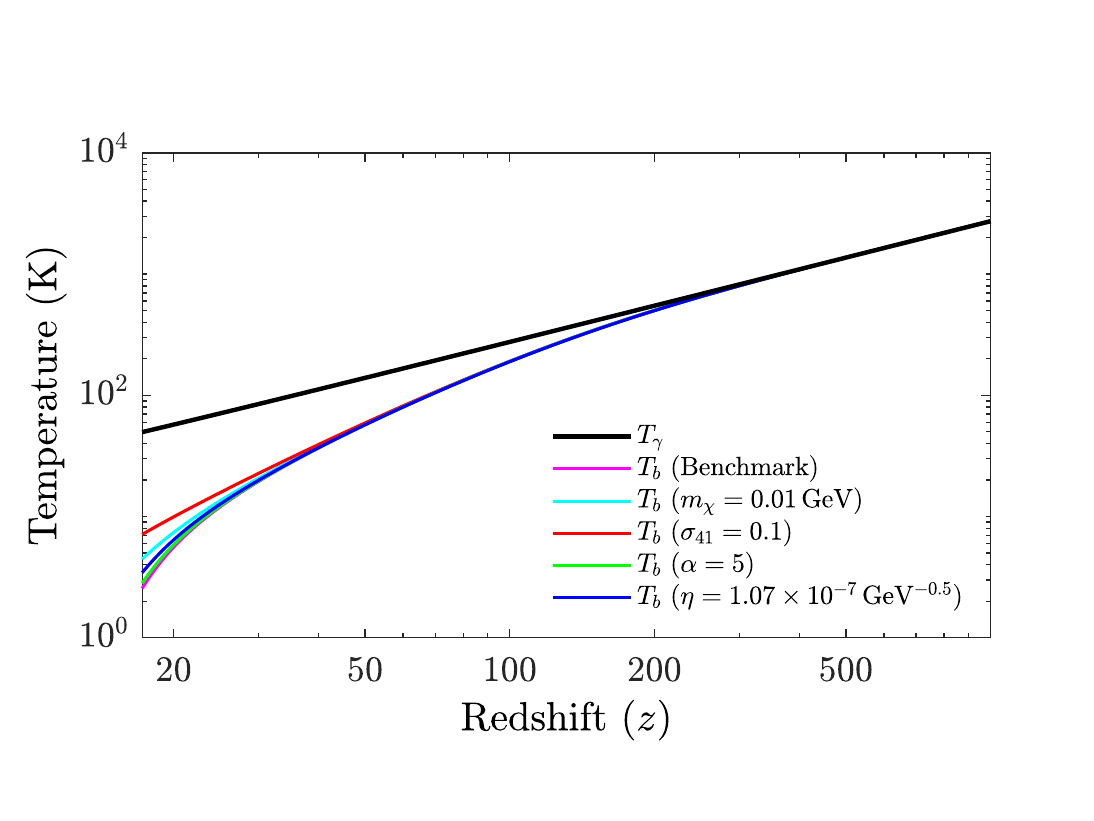}&
			\includegraphics[trim=0 40 0 0, clip, width=0.5\textwidth]{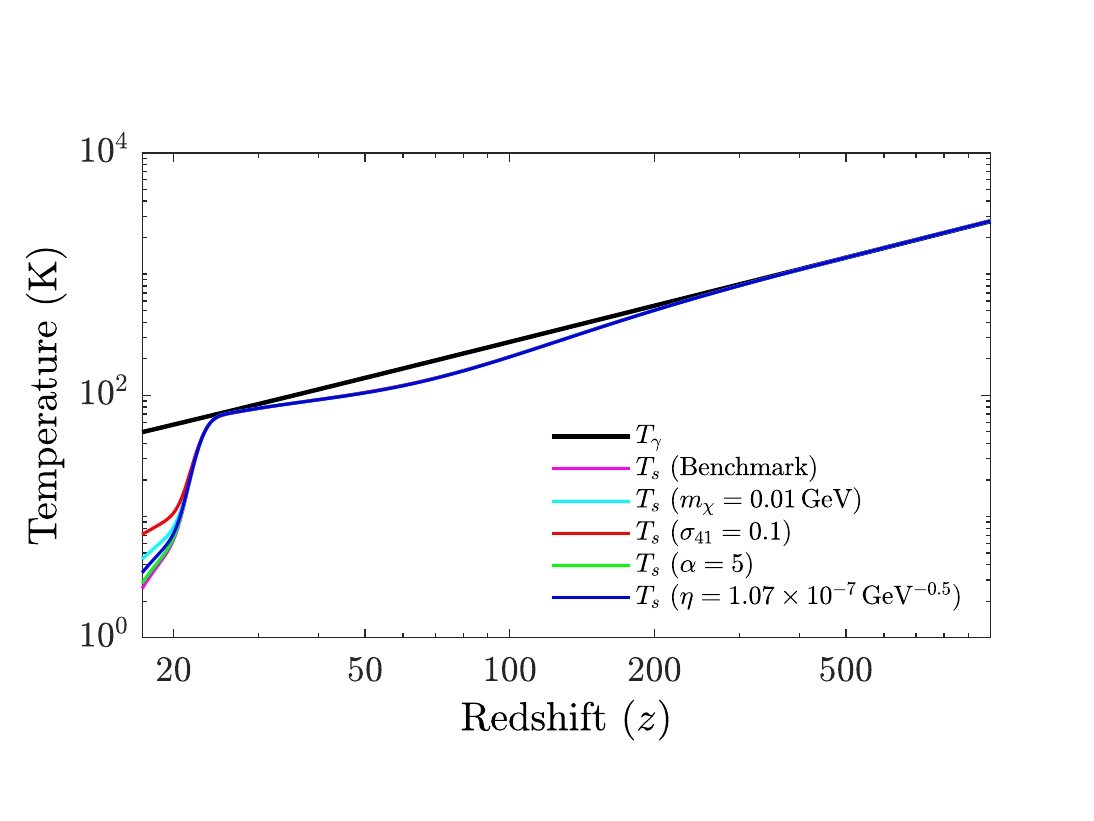}\\
            a & b\\
        \end{tabular}
        \begin{tabular}{cc}
			\includegraphics[trim=0 40 0 0, clip, width=0.5\textwidth]{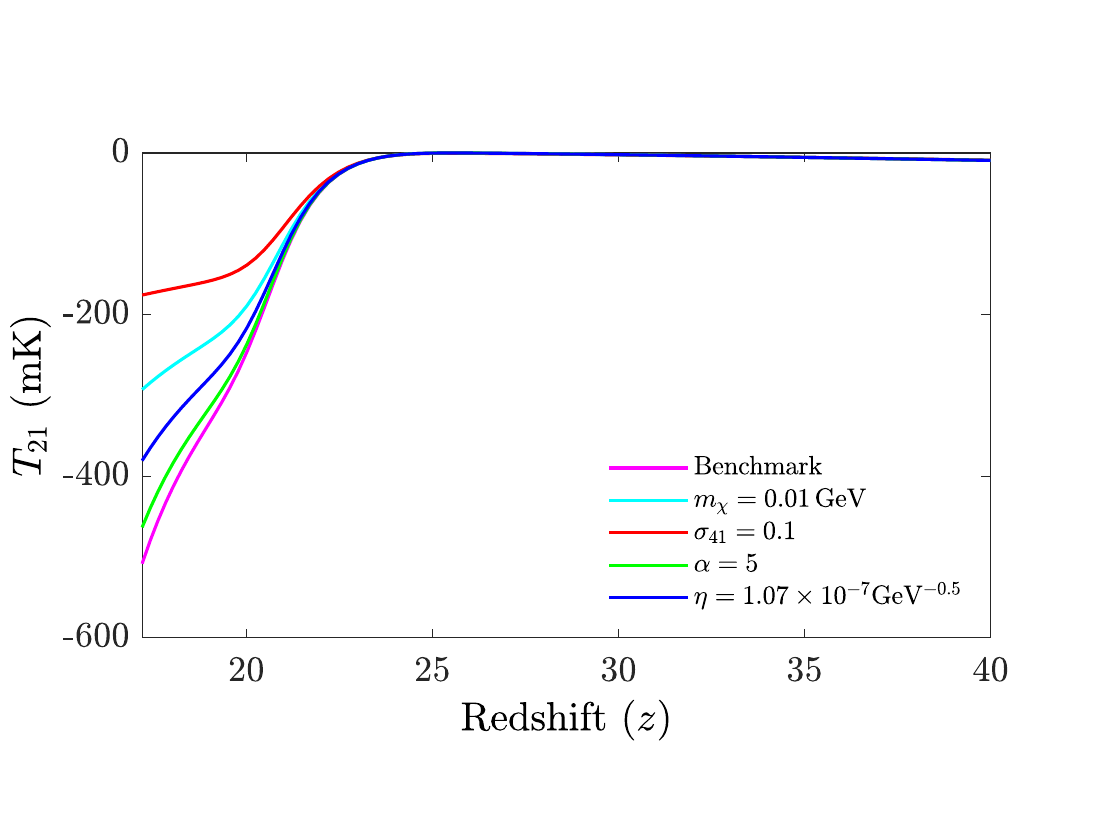}\\
			c\\
		\end{tabular}
		\caption{\label{fig:tspin} (a) Evolutions of baryon matter temperatures $T_b$ and (b) corresponding spin temperatures $T_s$ with redshift $z$ for different chosen values of model parameters. (c) Evolutions of brightness temperature $T_{21}$ with redshift $z$ for different chosen values of model parameters. See text for details.}
	\end{figure}
	
	\begin{table}[h]
	\begin{center}
		\begin{tabular}{C{0.15\textwidth}C{0.15\textwidth}C{0.15\textwidth}C{0.15\textwidth}}
			\hline \hline
			$m_{\chi}$ (GeV) & $\sigma_{41}$ & $\alpha$ & $\eta = \frac{\epsilon}{m_\chi^{-0.5}}$ ${\rm \left(GeV^{-0.5}\right)}$\\
			\hline
			$0.015$ & $0.25$ & $1$ & $1.05 \times 10^{-7}$\\
			\hline \hline
		\end{tabular}
		\caption{\label{tab:bench} Benchmark values of different parameters ($m_{\chi}$, $\sigma_{41}$, $\alpha$ and $\eta$) for the plots of Fig.~\ref{fig:tspin}.}
	\end{center}
	\end{table}
    The baryon temperature ($T_b$) and spin temperature ($T_s$) are obtained by solving numerically the coupled differential equations (Eqs.~\ref{eq:Vxb}, \ref{eq:xe}, \ref{txfinal} and \ref{tbfinal}) described in Sec.~\ref{sec:3}.
    In the present calculation, the benchmark values adopted for the parameters in the set of Eqs.~\ref{eq:Vxb}, \ref{eq:xe}, \ref{txfinal} and \ref{tbfinal}, we furnished Table.~\ref{tab:bench}. Note that $\eta$ is expressed in terms of $\epsilon$.
    The calculated values of $T_b$, $T_s$ and $T_{21}$ and their variation with redshift $z$ are shown in Fig.~\ref{fig:tspin}. The results are also compared with similar variations of $T_b$, $T_s$ and $T_{21}$ for different other parameter values (\textit{e.g,} $m_\chi$, $\sigma_{41}$). In Fig.~\ref{fig:tspin}(a), the magenta line represents the computed variation of $T_b$ with $z$ when the benchmark values from Table.~\ref{tab:bench} is used. Also shown a similar variation by changing each of the parameter values at a time while the other parameter values are kept fixed at the values given in Table.~\ref{tab:bench}. Similar plots for $T_s$ and $T_{21}$ are shown in Fig.~\ref{fig:tspin}(b) and Fig.~\ref{fig:tspin}(c) respectively. In all the plots of Fig.~\ref{fig:tspin}(a), Fig.~\ref{fig:tspin}(b) and Fig.~\ref{fig:tspin}(c), the variation of CMB (background) temperature $T_{\gamma}$ (black line) with $z$ are shown for reference. It is observed from Fig.~\ref{fig:tspin}(a) that the variation of $T_b$ with $z$ is not considerably sensitive to the change of the parameters such as the strength for cannibalism processes as also other parameters, \textit{e.g,} $m_\chi$ and $\sigma_{41}$. However, little variations (within $\sim 2^{\degree}$K - $7^{\degree}$K) are observed for $T_b$ values for $z \lesssim 30$. It can be also clearly seen from Fig.~\ref{fig:tspin}(a) that during reionization, the spin temperature for individual cases undergoes a rapid transition from $T_{\gamma}$ to $T_b$ for individual cases, which is due to the Wouthuysen-Field effect. Similar insensitivities are also seen for $T_s$ vs $z$ plot in Fig.~\ref{fig:tspin}(b). In this case, very little dependence on the parameters is obtained for $z \lesssim 22$. For example, at $z \sim 17.5$, where the value of $T_{21}$ changes from $T_{21} \sim -500$ mK to $\sim -410$ mK when the strength parameter $\eta$ for the process $2 \rightarrow 2$ is changed from the benchmark value of $\eta = 1.05 \times 10^{-7}{\rm{GeV}}^{-0.5}$ to $\eta = 1.07 \times 10^{-7}{\rm{GeV}}^{-0.5}$. It can also be seen from Fig.~\ref{fig:tspin}(c) that the change of other parameters affects the variation of $T_{21}$ more than those for the cases of $T_b$ and $T_s$.

	\begin{figure}
		\centering
		\begin{tabular}{cc}
			\includegraphics[width=0.5\textwidth]{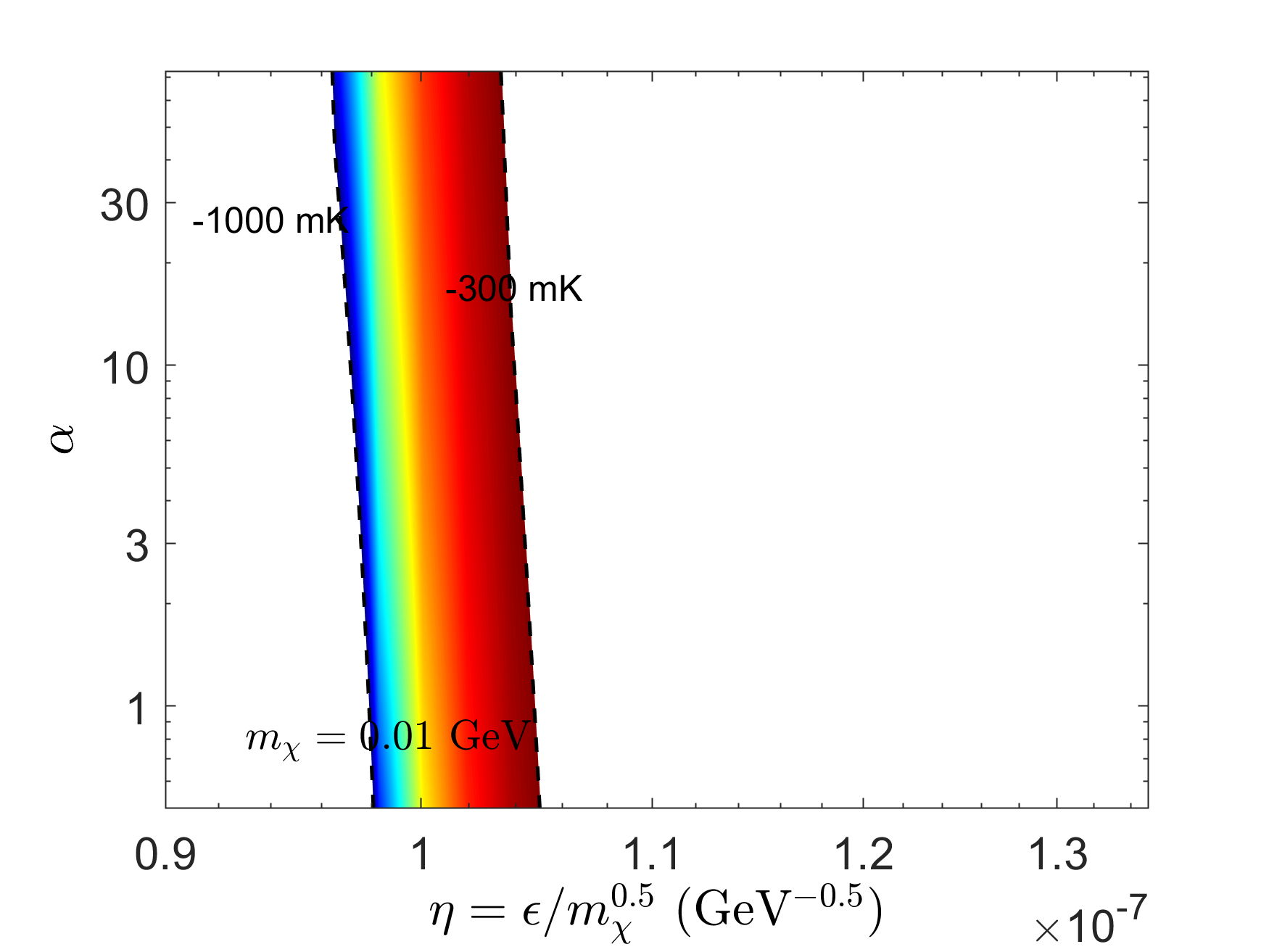}&
			\includegraphics[width=0.5\textwidth]{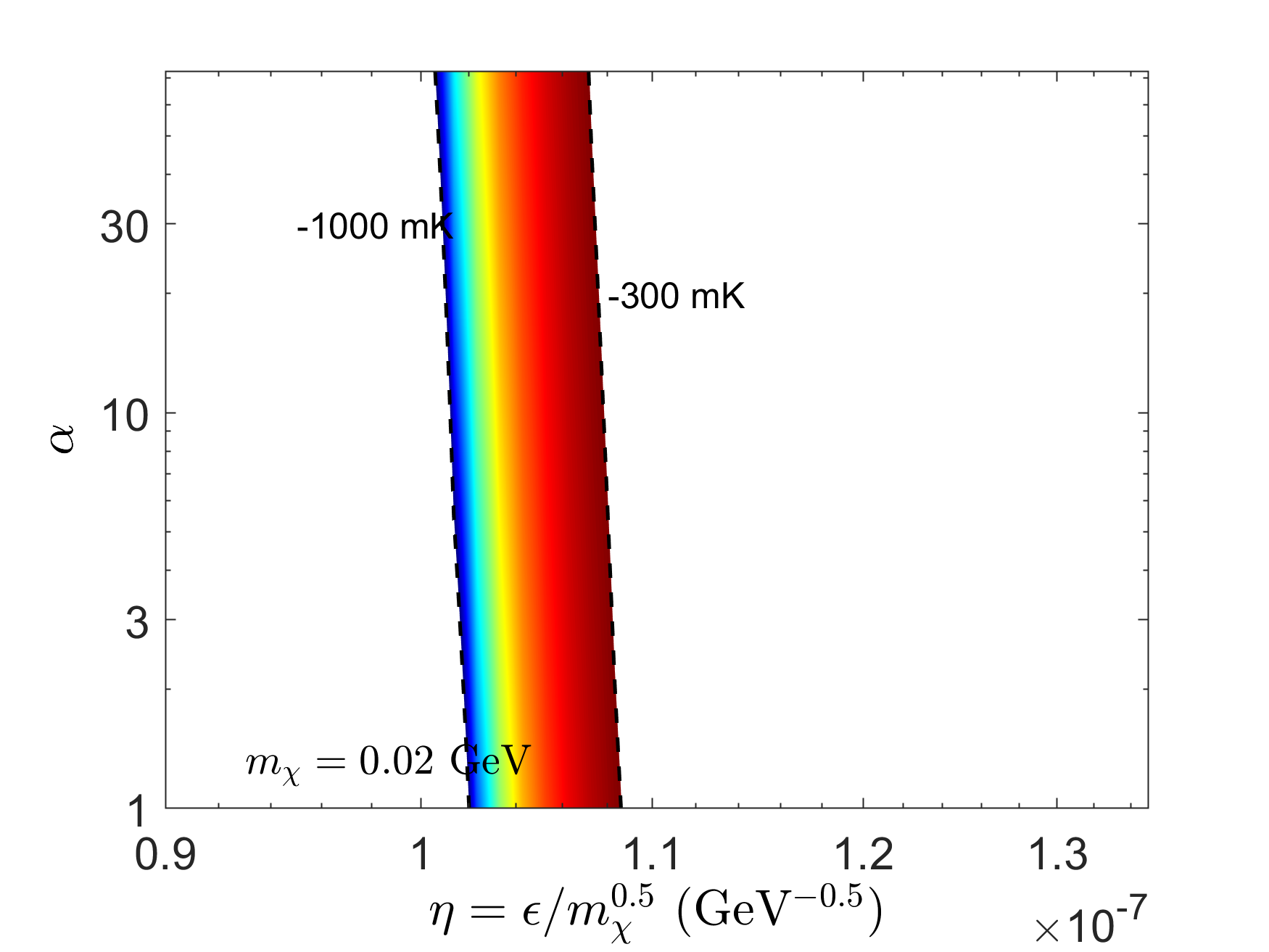}\\
			(a)&(b)\\
			\includegraphics[width=0.5\textwidth]{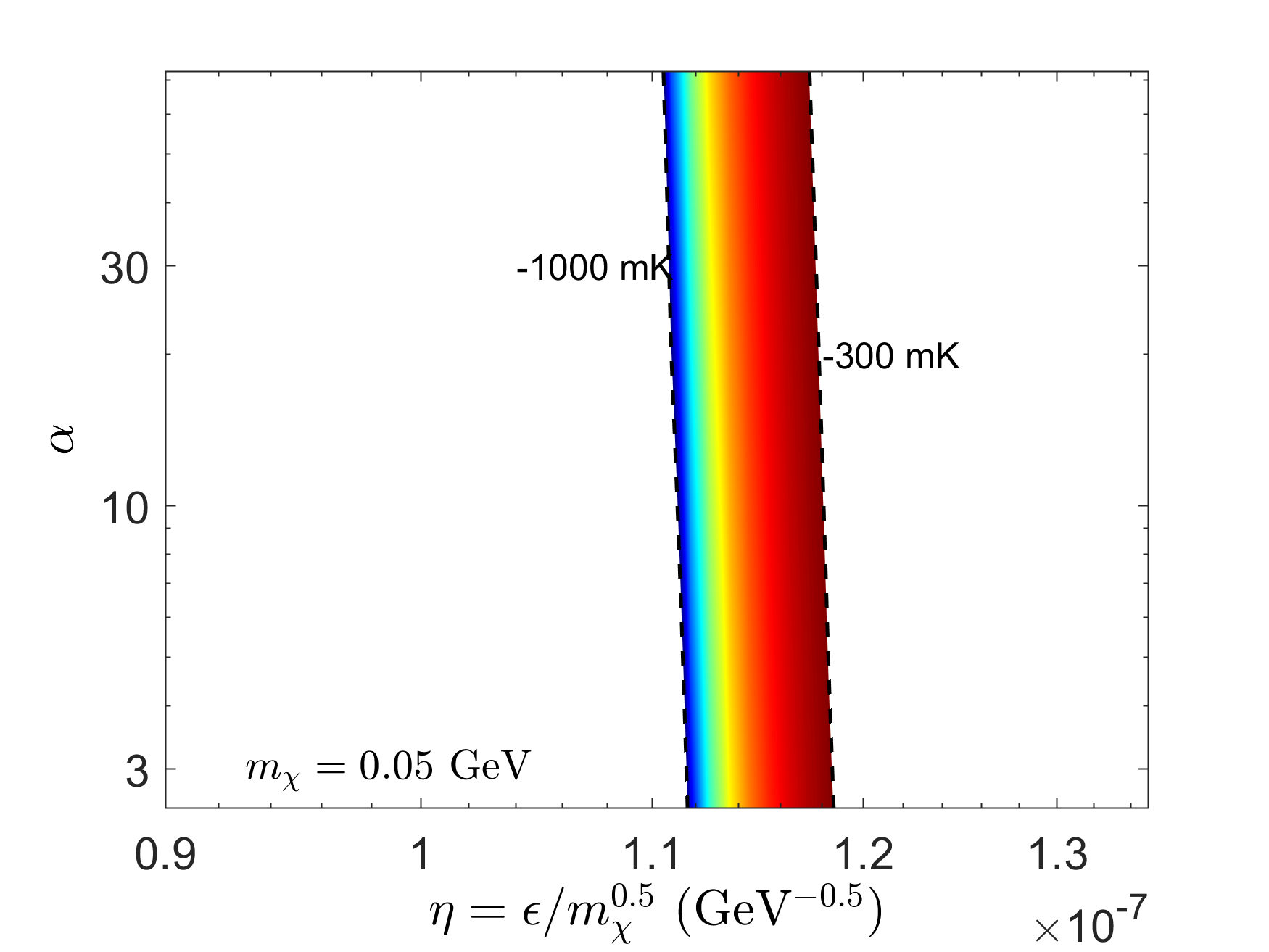}&
			\includegraphics[width=0.5\textwidth]{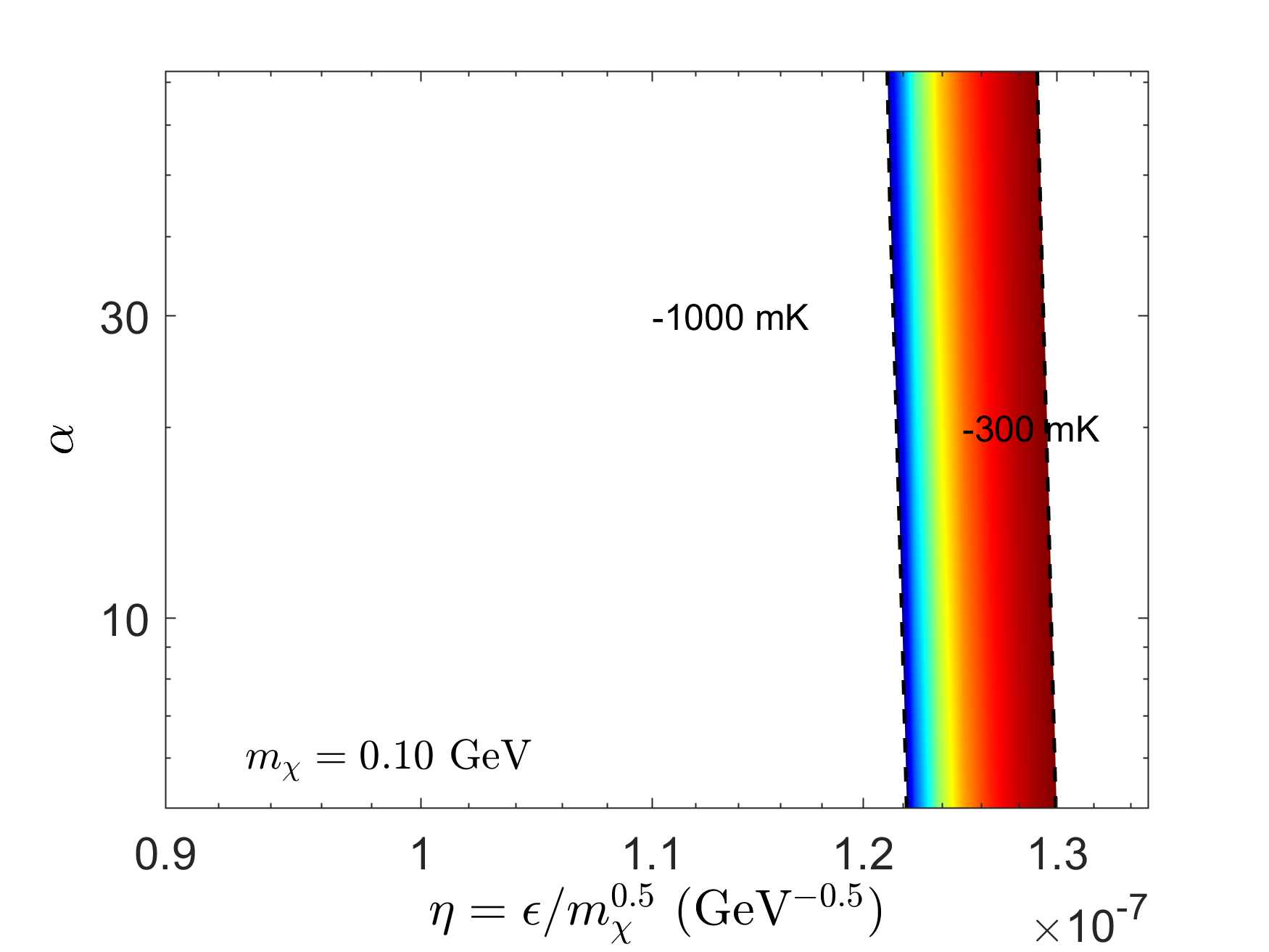}\\
			(c)&(d)\\
		\end{tabular}
		\begin{tabular}{c}
			\includegraphics[trim=0 10 0 220,clip, width=0.9\textwidth]{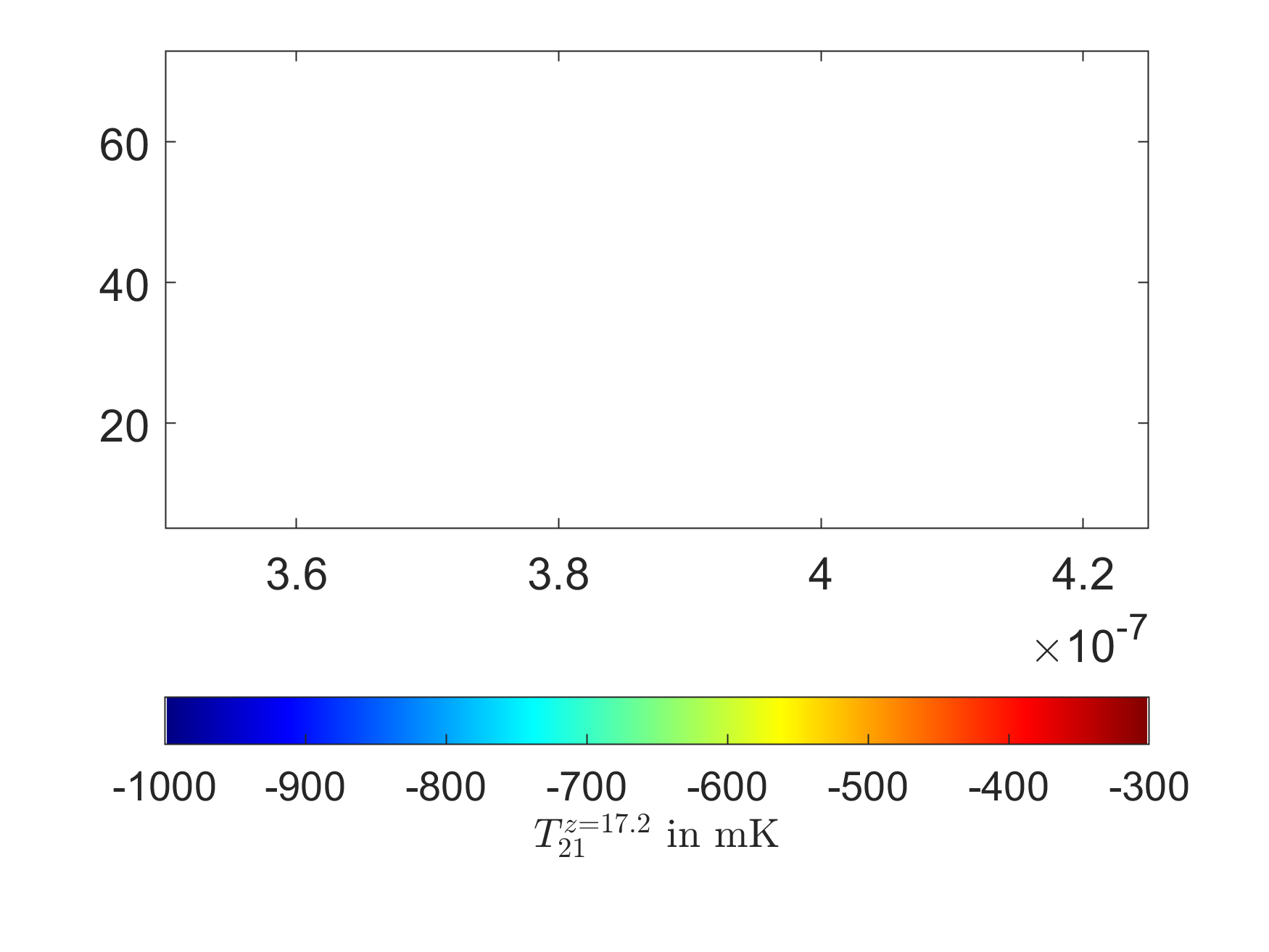}\\
		\end{tabular}
		\caption{\label{fig:eps_alpha} The allowed region in $\alpha$-$\eta$ parameter plane for (a) $m_{\chi}=0.01$ GeV, (b) $m_{\chi}=0.02$ GeV, (c) $m_{\chi}=0.05$ GeV and (c) $m_{\chi}=0.10$ GeV. In individual cases $\sigma_{41}=0.25$ is considered in all these plots. See text for details.}
	\end{figure}

	Now we extend our analysis to constrain, the parameters $\epsilon$, $\eta$ and $\alpha$ (various self-interacting parameters) using the EDGES observational results for brightness temperature $T_{21}$ at $z=17.2$ ($-300$ mK $\geq T_{21} \geq -1000$ mK). To this end, we introduce a parameter $T_{21}^{z=17.2}$, which denotes the brightness temperature $T_{21}$ at a particular redshift $z = 17.2$. In Fig.~\ref{fig:eps_alpha} the allowed zones in $\alpha$ - $\eta$ parameter plane are shown for different chosen values of dark matter mass $m_{\chi}$ namely $m_{\chi}=0.01$ GeV, 0.02 GeV, 0.05 GeV and 0.10 GeV. These are shown respectively in Fig.~\ref{fig:eps_alpha}(a), Fig.~\ref{fig:eps_alpha}(b), Fig.~\ref{fig:eps_alpha}(c) and Fig.~\ref{fig:eps_alpha}(d) respectively. In all the plots of Fig.~\ref{fig:eps_alpha}, different colours represent different values of $T_{21}^{z=17.2}$ and the colour index furnished at the bottom of Fig.~\ref{fig:eps_alpha}. The white regions in all plots of Fig.~\ref{fig:eps_alpha} represent the impermissible region according to the EDGES observational result. From Fig.~\ref{fig:eps_alpha} it can be seen that, as higher values of $m_{\chi}$ are taken into account, the value of $T_{21}^{z=17.2}$ decreases at individual points of the $\alpha$-$\eta$ plane. As a consequence, the allowed region shifted toward higher values of $\eta$. It can be seen that, the allowed zone extends to the entire allowed range of $\alpha$ (\emph{i.e.} $\alpha_{\rm min}$ and $\alpha_{\rm max}$) and is almost independent to $\alpha$. However for lower DM ($m_{\chi}$) a little variation is observed (see Fig.~\ref{fig:eps_alpha}(a) and Fig.~\ref{fig:eps_alpha}(b)). It is to be mentioned that, all plots of this figure (Fig.~\ref{fig:eps_alpha}) are plotted to keep $\sigma_{41}$ at $\sigma_{41}=0.25$.
	
	\begin{figure}
		\centering
		\includegraphics[width=\textwidth]{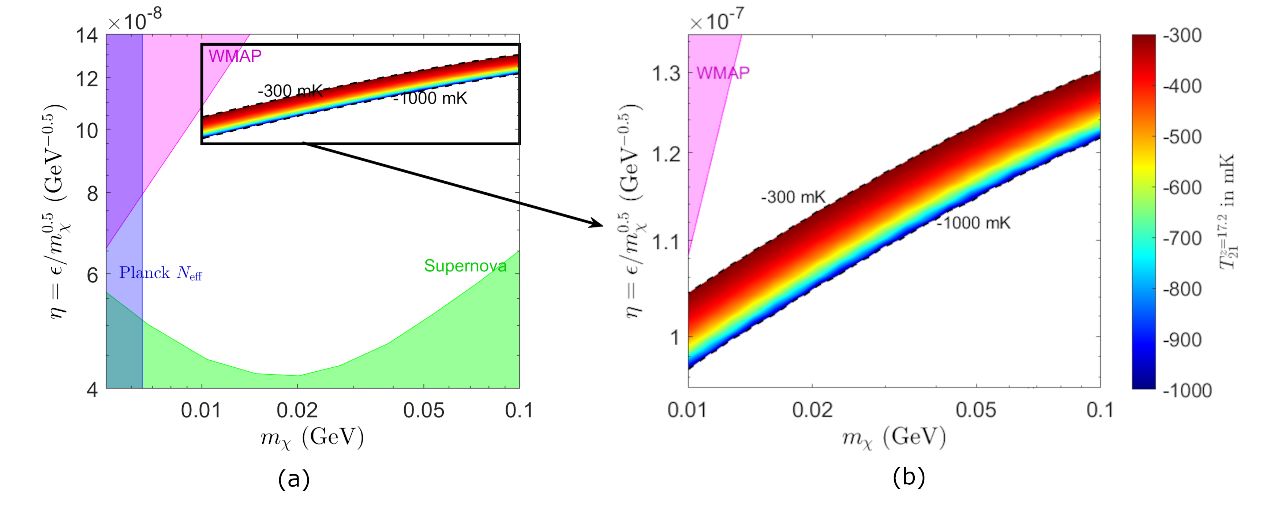}
		\caption{\label{fig:eps_mx} Similar allowed zone in $\eta$ - $m_{\chi}$ parameter plane.}
	\end{figure}

    From Fig.~\ref{fig:eps_alpha}, this can be observed that as $m_\chi$ increases, the higher values of $\eta$ is required to obtained a desired value of $T_{21}$. Clearly the allowed zone in Fig.~\ref{fig:eps_alpha} shifts towards right ($\eta$ increasing) as $m_\chi$ increases from Fig.~\ref{fig:eps_alpha}(a) - Fig.~\ref{fig:eps_alpha}(d). It is also evident from Fig.~\ref{fig:eps_alpha} that to achieve a specified variation in $T_{21}$ (within $-1000$mK and $-300$mK), the parameter $\alpha$ has to vary much more than that of $\eta$.  Therefore $T_{21}$ is more sensitive to $\eta$ than $\alpha$. These computations are made by adopting the value of $\sigma_{41}$ to be $0.25$.   
    In this case $\eta$ and $m_\chi$ are varied and the set of coupled differential equations are solved simultaneously with a fixed value for each $\sigma_{41}$ and $\alpha$.  In the present case, the value of $\sigma_{41}$ is chosen to be $\sigma_{41} = 0.25$. For fixing the parameter $\alpha$in these calculation, the range of $\alpha$ for a particular set of $\eta - m_\chi$ values and then the average value of $\alpha (=\dfrac{
    \alpha_{\rm{max}}+\alpha_{\rm{min}}}{2}$, where $\alpha_{\rm{max}}$ and $\alpha_{\rm{min}}$ are the maximum and minimum values of $\alpha$ for that range) is adopted for computing $T_{21}$ for that particular set of $\eta - m_\chi$ values. Also shown in Fig.~\ref{fig:eps_mx} are the comparisons with the bound obtained from Supernova, the WMAP result etc. The latter two bounds are adopted from Fig.~ 3 of Ref.~\cite{elder}. In case $\chi$ particle is trapped inside the SN core via the $\chi$ scattering on photons, it gives bound on $\epsilon$ \cite{elder}. WMAP bound is also adopted from the shown regime in Ref.~\cite{elder} and represented in Fig.~\ref{fig:eps_mx} in the present work. It can be seen from Fig.~\ref{fig:eps_mx} that the bound obtained from the present calculation using the EDGES 21-cm result also respects the bound from SN and WMAP. In Fig.~\ref{fig:eps_mx}, the constraint region from the present calculation is shown in the magnified version. The trapping of the dark matter $\chi$ provided via the process $\gamma \, \gamma \rightarrow \chi \, \chi$ at the Supernova (SN) is due to the elastic scattering of this dark matter $\chi$ on photons. This process involves the parameter $\epsilon$ and therefore lower bound on $\epsilon$ can be obtained \cite{elder} from the SN cooling via the energy loss rate on the process $\gamma \, \gamma \rightarrow \chi \, \chi$. This lower bound from Ref.~\cite{elder} is also reproduced in Fig.~\ref{fig:eps_mx} (in terms of $\eta$) for comparison. It can be observed from Fig.~\ref{fig:eps_mx}  that the $\eta - m_\chi$ allowed region in the present work, obtained with the observational limits of 21-cm line from the reionization epoch, also respects the SN constraint as well as PLANCK and WMAP constraints on $\epsilon$. It may be understood, from the zoomed version of the highlighted region (of Fig.~\ref{fig:eps_mx}(b)) in Fig.~\ref{fig:eps_mx} (a). Also shown in Fig.~\ref{fig:eps_mx} is the CMB constraints on DM annihilation into photons and the WMAP upper bound on $\eta(\epsilon)$ as adopted from Ref.~\cite{elder}.

	\begin{figure}
		\centering
		\includegraphics[trim=50 0 70 10, clip, width=0.5\textwidth]{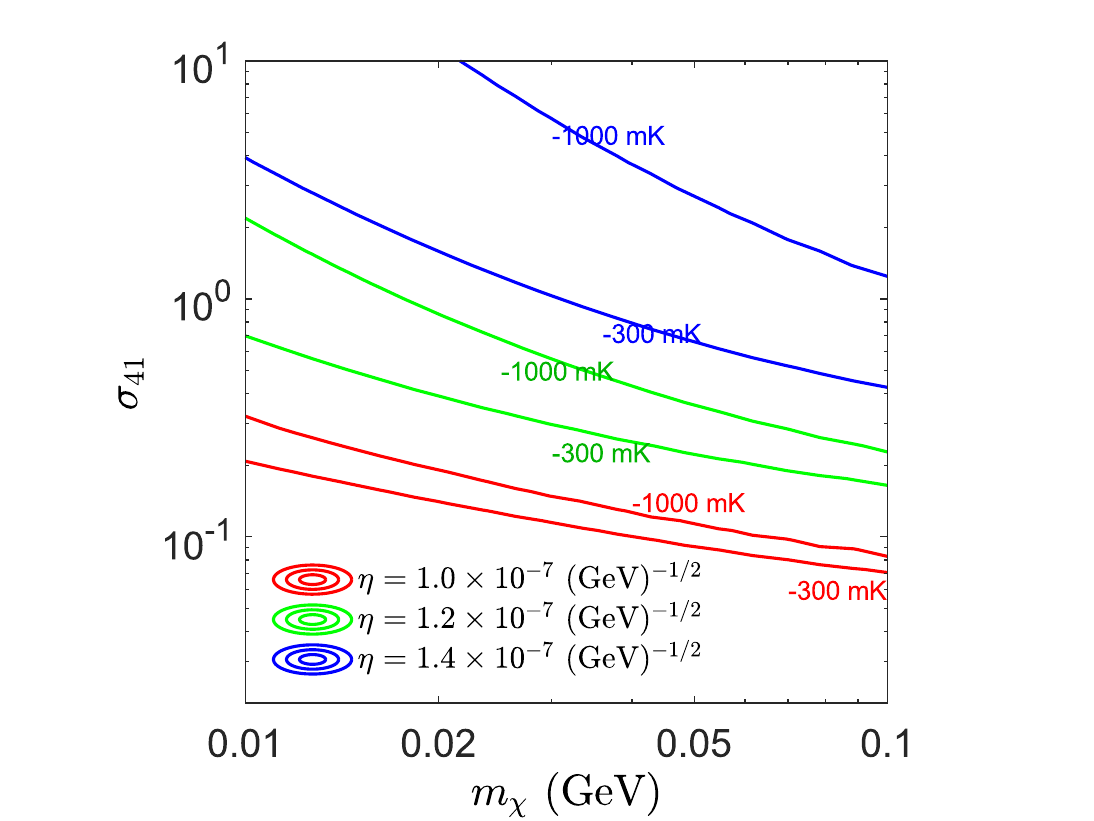}\
		\caption{\label{fig:mx_sig41}  The allowed regions in $\sigma_{41}$ - $m_{\chi}$ plane for different chosen values of $\eta$, along with several bounds estimated by different experiments.}
	\end{figure}
	The allowed region is also estimated in the $\sigma_{41}$ - $m_{\chi}$ space for three different chosen values of $\eta$. In Fig.~\ref{fig:mx_sig41}, the the allowed zones for three chosen values of $\eta$ namely, $\eta=1.0 \times 10^{-7}$ ${\rm GeV^{-1/2}}$, $\eta=1.2 \times 10^{-7}$ ${\rm GeV^{-1/2}}$ and $\eta=1.4 \times 10^{-7}$ ${\rm GeV^{-1/2}}$ within the DM mass range $0.01$ GeV$<m_{\chi}<0.1$ GeV are plotted along with several experimental bounds. In this analysis, the values of $\alpha$ at different values of $m_{\chi}$ are calculated by adopting the procedure, used to estimate the allowed region in Fig.~\ref{fig:eps_alpha} (\emph{i.e.} midpoint of the permissible range of $\alpha$ for different values of $m_{\chi}$). From this figure (Fig.~\ref{fig:mx_sig41}), one can see that the allowed region is very narrow at lower values of $\eta$. However, the region becomes wider and steeper at higher values of $\eta$. In addition, it is also observed that the allowed region corresponds to higher values of $\eta$ lies at higher $\sigma_{41}$ region. Also since the increase in $\eta$ (which corresponds to the DM self-interaction $\chi \,\chi \rightarrow \chi \, \chi$) may lead to injection of more heat ($\langle\sigma v\rangle \sim \dfrac{\epsilon^2}{m_\chi^2}$). On the other hand $\langle\sigma v\rangle_{\chi \,\chi \rightarrow \chi \, \chi}$ process increases as $m_\chi$ decreases. In order to compensate for this heating, the dark matter baryon scattering process should be increased which can also be seen from Fig.~\ref{fig:mx_sig41}.
	
	\begin{figure}
		\centering
		\includegraphics[scale=0.5]{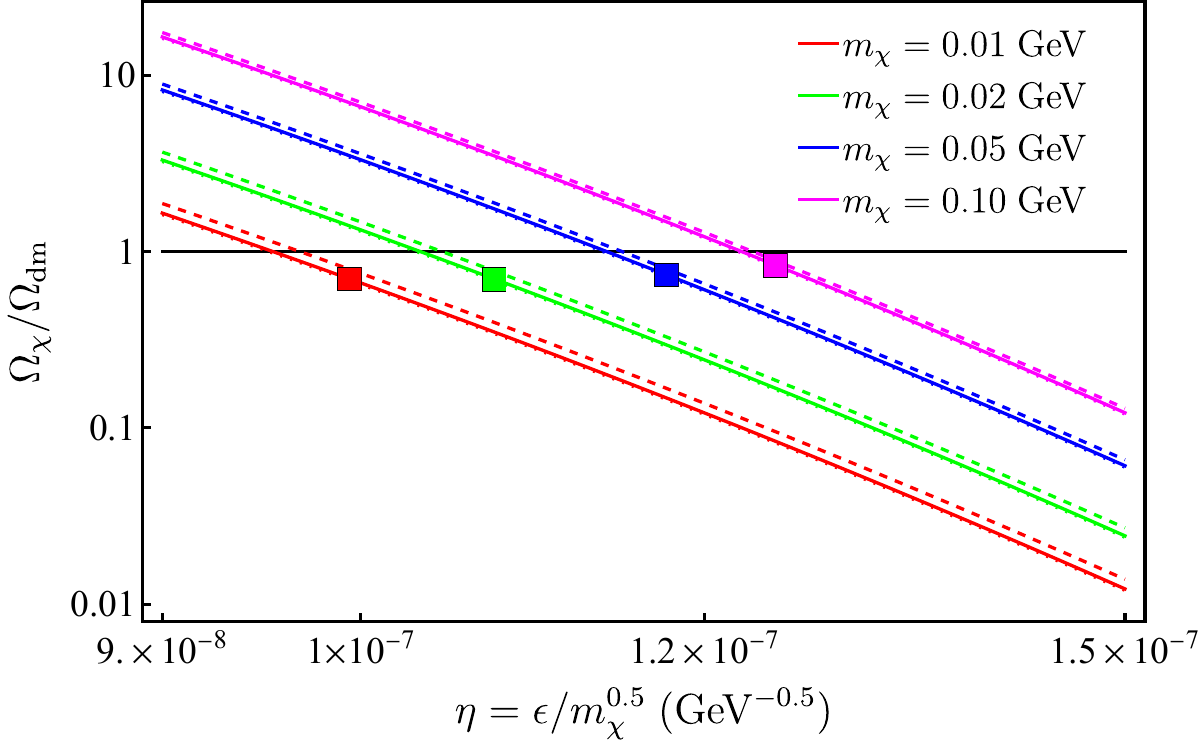}
		\caption{Variations of dark matter density $\Omega_\chi$ (normalized to the critical density of the Universe) for dark matter $\chi$ with $\eta$ in case of four different mass values of $\chi$. These are represented by solid colours lines. The dashed and dotted lines are for $\alpha = \alpha_{\rm{max}}$ and $\alpha = \alpha_{\rm{min}}(m_\chi)$ for each of the four cases ( the solid lines correspond to mean $\alpha$ values). The solid squares correspond to the EDGES result of brightness temperature $T_{21}^{z=17.2}=-500$ mK. See text for details. }
		\label{fig:omega}
	\end{figure}
	Finally we extend our analysis to estimate the values of density parameter for the ELDER type cold dark matter candidates $\Omega_{\chi}$ for different values of $\eta$. In Fig.~\ref{fig:omega}, the density parameter $\Omega_{\chi}$ is scaled as $\Omega_{\chi}/\Omega_{\rm dm}$, where the dark matter density parameter for the total dark matter contain of the Universe is denoted by $\Omega_{\rm dm}$ (adopted from the PLANCK \cite{planck} experimental result for $\Omega_{\rm DM}$). In this figure (Fig.~\ref{fig:omega}), the values of $\Omega_{\chi}/\Omega_{\rm dm}$ for different $\eta$ are plotted by the solid red line for the case of $m_{\chi}=0.01$ GeV. This line is estimated for $\alpha=\left(\alpha_{\rm max}+\alpha_{\rm min}(m_{\chi})\right)/2$ (as mentioned earlier), while the same for $\alpha=\alpha_{\rm max}$ and $\alpha=\alpha_{\rm min}(m_{\chi})$ are represented by the red dotted and the red dashed line respectively. Similar variations of $\Omega_\chi$ with $\eta$ for other three $m_\chi$ values are represented by sets of three lines with same colour for a fixed $m_\chi$. In each set the solid line of a particular colour is for $\alpha = (\alpha_{\rm{max}}+\alpha_{\rm{min}})/2 $ while the dashed and dotted lines of the same colour and set show the variations when $\alpha = \alpha_{\rm{max}}$ and $\alpha =\alpha_{\rm min}(m_\chi)$ respectively. In Fig.~\ref{fig:omega} the results for 
	$m_{\chi}=0.02$ GeV, $m_{\chi}=0.05$ GeV and $m_{\chi}=0.10$ GeV are plotted  with green, blue and magenta colour respectively. In addition, the values in the $\Omega_\chi/\Omega_{\rm DM} - \eta$ plane that represent the central value of EDGES experimental result  (which correspond to the brightness temperature $T_{21}^{z=17.2}=-500$ mK) for each of the four $m_\chi$ values considered in Fig.~\ref{fig:omega} are shown by solid square of the respective colours (corresponding to each $m_\chi$) in Fig.~\ref{fig:omega}. From this figure (Fig.~\ref{fig:omega}), it can be seen that the best fitted points for all these four chosen values of $m_{\chi}$ lie in the range $\Omega_{\rm DM} < 1$. This signifies that there are other dark matter particles in the Universe and thus there is a possibility of the existence of multi-component dark matter.

	\section{\label{sec:6}Summary and Discussions}
	The global 21-cm line of neutral hydrogen can be a useful probe to the reionization era following the ``Dark Ages''. This measure of the spectrum of the global 21-cm line namely the brightness temperature of the 21-cm line,  $T_{21}$, can be affected by, among other factors, the dark matter baryon interactions and other dark matter self-interaction processes. This includes dark matter-baryon elastic scattering and self-annihilation. In this work, the dark matter self-scattering and self-annihilation process are investigated in the light of the global 21-cm line signal from the era of (and leading to) the reionization  and the epoch of the first star formation. The process of dark matter-dark matter interaction that leads only to dark matter in the final product, can proceed in the channels that conserve the number of dark matter as also through the channels where the number of dark matter particles in the initial state are depleted at the final product. Thus while $``2\rightarrow2"$  process (2 dark matter particles scatter or annihilate to produce 2 dark matter particles at the final state) is a number conserving process, the possible process such as  $``3\rightarrow2"$  or $``4\rightarrow2"$ are called ``Cannibalism'' feature of dark matter interaction whereby the number of dark matter particles at the final end gets depleted.  In the present work we consider mainly $``2\rightarrow2"$  and $``3\rightarrow2"$ processes and investigate interaction strength of these processes in the light of brightness temperature  ($T_{21}$) results of 21-cm line from the era of cosmic dawn. This type of self-interacting DM is termed as ELDER dark matter. We find that the $``2\rightarrow2"$  process is more favoured than the cannibalism process $``3\rightarrow2"$. This can be concluded from the fact that to produce a certain change in $T_{21}$ value, the strength parameter $\alpha$ for $``3\rightarrow2"$ process has to be varied much more than the variation of the same for $``2\rightarrow2"$  process. Using the 21-cm signal from the era of cosmic dawn, the allowed regions for the $``2\rightarrow2"$  strength parameter and dark matter (DM) mass $m_\chi$  are also obtained in this work. It is seen that the $``2\rightarrow2"$  interaction strength normalised by DM mass increases with the increases of DM mass. The $``2\rightarrow2"$  interaction also seems to affect the DM-baryon elastic scattering cross-section. This is evident from the fact that the allowed region for ($\sigma_{41}-m_\chi$) space (with respect to EDGES $T_{21}$ results) moves towards higher $\sigma_{41}$ regime as the value of $\eta (=\dfrac{\epsilon}{m_\chi^{0.5}})$ increases. The area of the region also increases as higher value of $\eta$ is chosen. This is understandable from the fact that $``2\rightarrow2"$  interaction has a heating effect. In order to compensate for this heat by dark matter-baryon interaction so that the 21-cm spectrum cools down to the experimentally obtained results, higher values of $\sigma_{41}$ would be useful. It has also been found that the dark matter density $\Omega_{\chi}$ required to produced the EDGES results is always less than the total dark matter density $\Omega_{\rm DM}$ (given by PLANCK experiments) for wide range of values of $\eta$ and for different $m_\chi$ values (mass of dak matter $\chi$). This may be indicative of the presence of other dark matter components of different mass ranges in the Universe.
	
	This may be mentioned that, the $``3 \rightarrow 2 "$ process of the ELDER scenario implies that the  $``2 \rightarrow 2 "$ process has significant contribution in cosmological processes \cite{elder}. The cross-section for $``2 \rightarrow 2 "$ process is constraint as from Bullet Cluster and halo shapes observations, as $\dfrac{\sigma_{\chi \, \chi \rightarrow \chi \, \chi }}{m_\chi} \lesssim 1\, {\rm cm^2/g}$. This bound could be important for explaining the observed small-scale structure via N-body simulation results. The self-scattering of dark matter can thus play a measure role in the process of structure formation of the Universe.

	\section*{Acknowledgements}
	Two of the authors (S.B. and R.B.) wish to acknowledge the support received from St. Xavier’s College, Kolkata. R.B. also thanks the Women Scientist Scheme-A fellowship (SR/WOS-A/PM-49/2018), Department of Science and Technology (DST), Govt. of India, for providing financial support.
	
	\bibliographystyle{JHEP}
	\bibliography{ref}

\end{document}